 \documentclass[manuscript]{aastex}

\usepackage{color}

\shorttitle{High Precision CTE of SiC-100}
\shortauthors{Enya et al.}

\begin{document}


\title{
High Precision CTE-Measurement of SiC-100
for Cryogenic Space-Telescopes
}



\author{
K. Enya\altaffilmark{1}, N. Yamada\altaffilmark{2}, 
T. Onaka\altaffilmark{3}, 
T. Nakagawa\altaffilmark{1}, H. Kaneda\altaffilmark{1},
M. Hirabayashi\altaffilmark{4}, \\
Y. Toulemont\altaffilmark{5}, D. Castel\altaffilmark{5},
Y. Kanai\altaffilmark{6}, and N. Fujishiro\altaffilmark{6}
}

\email{enya@ir.isas.jaxa.jp}

\altaffiltext{1}{
Institute of Space and Astronautical Science, 
Japan Aerospace Exploration Agency, 
3-1-1 Yoshinodai, 
Sagamihara, Kanagawa 229-8510, Japan
}

\altaffiltext{2}{
National Metrology Institute of Japan, 
Advanced Industrial Science and Technology,
3 Tsukuba Central, 
Tsukuba, Ibaraki 305-8563, Japan
}

\altaffiltext{3}{
Department of Astronomy,
Graduate School of Science, University of Tokyo,
7-3-1 Hongo, 
Bunkyo-ku, Tokyo 113-0033, Japan
}

\altaffiltext{4}{
Sumitomo Heavy Industries, Ltd., 
 5-2 Niihama Works, 
Soubiraki-cho, Niihama, Ehime 792-9599, Japan
}

\altaffiltext{5}{
Astrium Satellites, Earth Observation, 
Navigation \& Science\,(F),
31 av des Cosmonautes, 
31042 Toulouse Cedex 4, 
France 
}

\altaffiltext{6}{
Genesia Corporation,
Mitaka Sangyo Plaza 601,
3-38-4 Shimorenjyaku, 
Mitaka, Tokyo 181-0013, Japan
}



\begin{abstract}
We present the results of high precision measurements of the 
thermal expansion of the sintered SiC, SiC-100, intended for use in
cryogenic space-telescopes,
in which minimization of thermal deformation of the mirror
is critical and precise information of the thermal expansion
is needed for the telescope design.   
The temperature range of the measurements extends 
from room temperature down to $\sim$\,10\,K. 
Three samples, \#1, \#2, and \#3 were manufactured from 
blocks of SiC produced in different lots. 
The thermal expansion of the samples was measured with a 
cryogenic dilatometer, consisting of a laser interferometer,
a cryostat, and a mechanical cooler. 
The typical thermal expansion curve is presented using the 
8th order polynomial of the temperature.
For the three samples, 
the coefficients of thermal expansion\,(CTE),
$\overline{\alpha}_{\#1}$,  
$\overline{\alpha}_{\#2}$, 
and $\overline{\alpha}_{\#3}$ 
were derived for temperatures between 293\,K and 10\,K.
The average and the dispersion\,(1\,$\sigma$ rms) 
of these three CTEs 
are 0.816 and  0.002\,($\times\,10^{-6}$/K), respectively. 
No significant difference was detected 
in the CTE of the three samples from the different lots. 
Neither inhomogeneity nor anisotropy of the CTE was observed.
Based on the obtained CTE dispersion, 
we performed an finite-element-method (FEM) 
analysis of the thermal deformation 
of a 3.5\,m diameter cryogenic mirror made of six SiC-100 segments. 
It was shown that the present CTE measurement has a sufficient 
accuracy well enough for the design of the 3.5\,m cryogenic infrared 
telescope mission, 
the Space Infrared telescope for Cosmology and Astrophysics\,(SPICA).

\end{abstract}






\section{Introduction}

Development of cryogenic light-weight mirrors is a key 
technology for 
infrared astronomical space-telescope missions, 
which have large advantages 
owing to being free from the turbulence, 
thermal background, and absorption caused by atmosphere.
The light-weight mirror technology is essential
to bring large mirrors into space that enable 
a high sensitivity and a high spatial resolution. 
The infrared sensitivity of the space-telescope 
is vastly improved by reduction of the thermal background 
by cooling the telescope to cryogenic temperatures.
Thus infrared space telescopes badly need cooled 
light-weight mirrors
with a sufficient optical quality.

Silicon-carbide (SiC) is one of the most promising materials for 
space telescopes because of its
high ratio of stiffness to density. 
Japanese space mission for infrared astronomy, {\it AKARI},
carries a 68.5 cm aperture telescope
whose mirrors are made of  sandwich type 
SiC \citep{murakami2004,kaneda2005}.
The entire {\it AKARI}  telescope system is cooled
down to 6\,K by helium gas vaporizing from liquid helium. 
{\it AKARI} was launched in February 2006 and 
the telescope system performance has been confirmed 
to be as expected from pre-launch ground tests 
{\citep{kaneda2007}.}
The Herschel Space Observatory,
a submillimeter satellite mission 
by the European Space Agency \citep{pilbratt2004},
employs mirrors of sintered SiC\,(SiC-100)
provided by Boostec Industries 
and EADS-Astrium \citep{breysse}.
The 3.5m diameter primary mirror of the
Herschel telescope is made of 12 segments brazed together. 
The Herschel Space Observatory
will be launched in 2008 to make observations in
60--670\,$\mu$m,  
whilst the telescope will be kept
to $\sim$\,80\,K by passive cooling.
SiC-100 is one of the most frequently used SiCs for space optics. 
It has been used for the telescopes of ALADIN, GAIA, ROCSAT and other 
missions\,(Breysse et al. 2004 and the references therein).

The Space Infrared telescope for Cosmology and 
Astrophysics\,(SPICA) is the next generation 
mission for infrared astronomy, 
planned by the 
Japan Aerospace Exploration Agency \citep{nakagawa2004, onaka2005}. 
The SPICA telescope is required to have a 3.5\,m diameter aperture 
and will be cooled down to 4.5\,K by the combination of radiative 
cooling and mechanical coolers \citep{sugita}. 
SPICA is planned to be launched
in the middle of the 2010s and execute infrared 
observations in 5--200\,$\mu$m.
SiC-100 is one of the promising candidate 
materials for the mirrors and structures of 
the SPICA telescope, 
whilst carbon-fiber 
reinforced silicon-carbide is another candidate now being investigated
\citep{ozaki2004,enya2006a,enya2004}.
Monolithic primary mirrors can be manufactured by the 
joint segment technology, similar to the primary mirror for 
the Herschel telescope.
The requirement for the surface figure accuracy of the
SPICA primary mirror is, however,  better than 0.06\,$\mu$m rms. 
This requirement is $\sim 20$ times more severe than the 
Herschel Space Observatory
because of the
difference in the targeted wavelength range.

In the development of cryogenic space-telescopes, 
it is important to suppress the thermal deformation 
of the mirror caused by cooling to
satisfy the requirement for the surface figure 
accuracy \citep{kaneda2003,kaneda2005}.
Therefore, the study of the coefficients of  thermal expansion\,(CTE) 
of the material used for the mirror is important. 
The CTE data of the mirror material are indispensable 
for the design of cryogenic space-telescopes
because the actual telescope mirror needs to accommodate
complicated support structures consisting of 
materials different from the mirror.  
If the CTE measurement of test
pieces of the mirror material has a sufficient accuracy, 
it will enable us to predict the thermal deformation
of a segmented mirror caused by the dispersion in the 
CTE of each segment.
The most direct and highly sensitive test of the
thermal deformation is an interferometer measurement
of the actual mirror at cold temperatures.
The CTE data are useful to interpret the result of 
direct measurements of the mirror
and investigate the origin of the observed deformation.

However, measurements of the CTE of SiC and 
its dispersion have not so far been performed with a sufficient 
accuracy and 
{for} a wide temperature range.
Especially, little data are available at temperatures lower 
than 77\,K, which can be realized by only liquid nitrogen cooling.
Prior to this work, available CTE data of SiC-100 were limited
at temperatures higher than $\sim$\,77\,K \citep{toulemont}. 
Pepi \& Altshuler\,(1995) presented the CTE data of reaction
bonded optical grade\,(RBO) SiC down to 4\,K based on measurements
with samples made from one block of the RBO SiC.
The CTE of the new-technology {SiC} (NT-SiC), developed high-strength 
reaction-sintered SiC, has been reported down to 20\,K  
{for} one sample by Suyama et al.\,(2005).

In this work, 
we present the results of high precision CTE measurements 
of SiC-100 down to cryogenic temperatures
for {samples from three different} lots.
We set two major goals for the present work: 
one is to provide the typical CTE of 
SiC-100 for the development of the SPICA telescope, 
and the other is to estimate 
the thermal deformation of segmented mirrors
from the measured CTEs.

\section{Experiment}

\subsection{Sample}

Figure\,1 shows three samples\,(\#1, \#2, and \#3) of SiC-100 
measured in this work. 
All of the samples were manufactured by Boostec Industries and 
EADS-Astrium \citep{breysse}.
Each of the samples was extracted
from blocks of SiC  produced in different lots. 
The locations of each sample in the SiC blocks are arbitrary.
The samples are of a rectangular
parallelepiped 
shape with the dimension of
$20.00_{-0.00}^{+0.05}$\,mm\,$\times 20.00_{-0.00}^{+0.05}$\,mm\,$\times
6.0_{-0.0}^{+0.1}$ mm.
Flatness, parallelism, and roughness of the 
$20.00$\,mm\,$\times6.00$\,mm
surfaces are important for the measurement of this work: 
The flatness of these surfaces is $\leq\,\lambda/10$ rms,
where $\lambda$ is the wavelength of the He-Ne laser, 632.8\,nm.
The parallelism of {the} opposing 
$20.00$\,mm\,$\times6.00$\,mm
surfaces is less than 2$^{\prime\prime}$, 
whilst the parallelism of {the} opposing 
$20.00$\,mm\,$\times 20.00$\,mm 
surfaces is less than 1.0$^{\circ}$. 
All of the  $20.00$\,mm $\,\times 6.00$\,mm 
surfaces are polished to an optical grade and the 
surface roughness finally achieved is less than 3 nm rms. 
Owing to the polished surface, 
the directly reflected laser light by the sample
was used for the measurement.
As the result, the measurement was free from the
uncertainty due to any additional mirrors or 
coating, which would be needed 
if the direct reflected light could not be used.

\subsection{Measurement}

The measurement of the thermal expansion 
was carried out with the laser interferometric 
dilatometer system for low temperature
developed in the National Metrology Institute of Japan, 
Advance Industrial Science and 
Technology\,(Yamada \& Okaji 2000, hereafter the paper--I;
Okaji \& Yamada 1997; Okaji et al. 1997). 
The system consists of the cryostat, the cryogenic 
mechanical refrigerator of
the GM cycle\,(V204SC) by Daikin Industries Ltd., 
and the interferometer utilizing acousto-optical 
modulators and the stabilized 
He-Ne laser system\,(05STP905) by Melles Griot.
The cooling is performed with the refrigerator 
alone and no cryogen is needed. 
The minimum temperature achieved in this work is about 10\,K. 
The space around the cold stage of the cryostat, 
containing the installed sample, 
is filled with 130\,Pa helium gas and sealed off
at room temperature before cooling
to ensure thermal uniformity. 
The configuration of the whole cryostat and
the sample installation into the cold stage of the 
cryostat are shown in figures\,1 and 2 of the paper--I, respectively. 
The change of length of the sample, $\Delta L$,
is measured with the double path type laser interferometer {of}
the optical heterodyne method 
with the digital lock-in amplifier\,(Model SR-850) 
from Stanford Research System. 
Details of this system are given in the paper--I.

We made totally 6 measurements of the CTEs of SiC-100.  
In each measurement,  the sample was initially cooled down to
$\sim 10$\,K  and then the proportional integral derivative (PID) 
temperature control was applied. 
After the the temperature had been stabilized, 
the temperature and the change of the sample length $\Delta L$
were measured.
The temperature stability during the
measurement was less than 0.02\,K per hour. 
This process was repeated at approximately 16
temperatures
of roughly equal intervals up to room temperature. 
One dataset of the temperature
vs.  $\Delta L$
was obtained for one cooling cycle.  To compensate the systematic
uncertainty, each measurement was repeated by rotating the sample 
by 90$^{\circ}$ as described in figure\,2 of the paper--I.
We measured CTEs of two orthogonal directions (A and B directions as shown
in figure\,1) of each sample.  The directions A and B were arbitrary chosen.

\section{Result and discussion}

\subsection{Typical thermal expansion}
The results of the measurement of the thermal expansion
are presented in figure\,2.
A fit with the 8th order polynomial 
is applied to each of the six datasets
and the thermal expansion (contraction) 
$\Delta L/L$ is set as 0 at 293\,K for each of the six curves{.}
The six datasets are plotted in figure\,2\,(a). 
We present the curve derived from the fit with all of the
six datasets as the typical thermal contraction of SiC-100.
It is shown by the solid line in figure\,2\,(a).
The coefficients of the 8th order polynomial, 
$\Delta L/L = \sum_{i = 0}^8 a_i T^i$, are presented in table\,1.
Figure\,2\,(b) shows the residual dispersion of 
the data after subtracting the fit curve. 
The shape of the curve in figure\,2\,(a) 
is roughly compatible
with those of the RBO SiC \citep{pepi1995} 
and NT-SiC \citep{suyama2005},
though slightly negative CTEs observed in the
RBO SiC and NT-SiC observed at temperatures less than 50\,K
are not seen in the present measurements for SiC-100.
Because of the uncertainties in the other measurements, it is difficult
to further investigate the origins of the differences at present.

For each of the six curve{s}, the average
$\Delta L/L$ per temperature between 293\,K and 10\,K 
is derived and summarized in table\,2.
The average of the six values and their
dispersion\,(1\,$\sigma$) are 
0.816 and 0.005\,($\times\,10^{-6}$/K), respectively.
The  dispersion is smaller than
the previous upper-limit obtained 
with high-purity single crystal
{silicon in paper--I,
0.01\,($\times\,10^{-6}$/K), indicating that the present
measurements have reached the limit set by the instrument.

Pepi \& Altshuler\,(1995)
have shown the 1\,$\sigma$ dispersion of 0.04\,($\times\,10^{-6}$/K)
for their measurements.  
Karlmann et al.\,(2006)
presented the repeatability 
of the CTE measurement to be 0.004\,($\times\,10^{-6}$/K)
from 35\,K to 305\,K
of single crystal silicon by the
interferometer based cryogenic dilatometer.
{Comparing with Karlmann et al.\,(2006),}
our measurement reached lower temperature{s}.
Thus the present dispersion is concluded to be well below the 
measurement uncertainty and we do not detect any significant variations
in the CTEs of the present 6 measurements.  We do not detect either any
differences in the CTEs in {different} 
directions of the same sample
or any differences in the samples extracted from different lots.
Thus SiC-100 we have measured is homogeneous and isotropic within 
the present measurement uncertainty.
Finally, we average two  $\overline{\alpha}$
in directions A and B of each of three samples 
to obtain
$\overline{\alpha}_{\#1} =  0.8145$,  
$\overline{\alpha}_{\#2} = 0.8160$, 
and $\overline{\alpha}_{\#3} = 0.8185$\,($\times\,10^{-6}$/K).
The average and the dispersion\,(1\,$\sigma$ rms) 
of these three CTEs 
are 0.816 and  0.002\,($\times\,10^{-6}$/K), respectively.

\subsection{Alternative measurements of the dispersion of CTE}

Differential tests of the CTE dispersion 
is another strong tool to investigate the variations in the CTEs.
The quite small dispersion in the CTEs of SiC-100
has been confirmed by the
systematic check made by Astrium in the manufacturing process
of the mirror segments of the Herschel Space Observatory.  
All the 12 segments of the Herschel primary mirror came from 
different SiC powder batches. 
The measurement of the
bending deflection was made for a brazed couple of two thin SiC samples coming
from different batches of SiC. 
The couples were placed in a vacuum chamber and 
the thermal deformation of the surface figure 
of the samples was measured by a interferometer
between  room temperature and 150K. 
The deformation data is directly linked to the curvature 
of the bending deflection and to the difference 
of the CTE between the two samples.

In order to validate the sufficient
homogeneity of the SiC material, 
two  kinds of verification on the material have been performed:
One of the tests was of homogeneity inside one spare segment. 
For this test, samples were cut out from one spare sintered segment
{at}
several locations (along radial, tangent, and thickness directions). 
The other tests was of homogeneity between samples belonging 
to {different} flight segments. 
For this test, as it is was not possible to take {samples}
from the flight segments after sintering, the samples were 
cut out from {different} segments at their green body stage, 
before the sintering of the segments. 
Those samples were taken from  arbitrary locations of the segments 
(different orientations and locations were therefore 
present in the test samples).
Those samples were made similar to the segments themselves by taking 
care of sintering them in the same run as the associated segments. 
By this way, it was possible to reproduce 
the differential CTE characteristics.
In the  telescope manufacturing process, 12
samples of the 12 segments were brazed on the reference samples and tested
at 150K. 
The results indicate the dispersion (1\,$\sigma$ rms) 
to be smaller than $0.0025$\,($\times\,10^{-6}$/K).

The agreement of the dispersion of the CTE derived by 
two different methods signifies that 
both of the measurements are reliable and the uniformity 
of the CTE of the SiC-100 is well confirmed.    
The direct measurement of the CTE of this work
and the differential measurement are complementary: 
the direct measurement provides 
absolute CTE data down to \,10\,K with high accuracy, which is
indispensable to design the space telescope including the surrounding
structures, 
whilst the differential test checks
the uniformity of the CTE for a large number of the samples.

\subsection{Simulation of the mirror deformation}

{
It is fruitful to relate the accuracy of the CTE measurement
with corresponding thermal deformation of the mirror.
}
In this section the thermal deformation 
of the segmented mirror is estimated 
on the basis of the measured dispersion of the CTE values.
To examine this issue, 
we perform a case study by using a simple model 
of the finite-element-method (FEM) analysis. 
All of the simulated thermal deformation 
is derived for the case
of cooling down from 293\,K to 10\,K. 
Figure\,3 shows the model used in the FEM  analysis.
One mirror with a center hole
is constructed from six segments
with the rib structure as shown in figures\,3\,(a) and b as 
a light-weight mirror design. 
In this model, it was assumed that the mirror surface 
is flat for simplicity.
Figures\,3\,(c) and d show a three-dimensional view and  
the geometry of one  segment, respectively.
The diameter of the whole mirror is 3.5\,m, 
equating to the designed diameter
of the primary mirror of the SPICA telescope and
the Herschel Space Observatory. 
The thicknesses of the rib structure and
mirror surface are 3 mm. 
The Young's modulus and Poisson's ratio of 
SiC-100 {at} room temperature are reported to be 420\,GPa 
and 0.17{,} respectively
\citep{breysse,toulemont}.
We use these values in the simulation of the thermal deformation
since the temperature dependence 
of these quantities is usually small and not available at present. 
In the FEM analysis, the rotationally symmetric axis of 
the mirror is set along the z axis of the Cartesian {coordinates.} 
The constraints of the model are shown in figure\,3\,(b):
Three points on the rib of the mirror, indicated by triangles, 
are constrained on the lines in the x-y plane 
as shown by the arrows in figure\,3\,(b).
These points are free along the radial
direction within the constraint lines. 
Therefore, this constraints cause no inner stress 
{in} the mirror in the simulation of the cooling 
and {thermal} deformation.

The results of the simulation 
are  presented in figure\,4.
Figure\,4\,(a) shows the thermal deformation 
toward the z-direction of the mirror surface obtained
from the FEM analysis, in which
$\overline{\alpha}_{\#1}$, 
$\overline{\alpha}_{\#2}$, and
$\overline{\alpha}_{\#3}$ 
are given for the segment 
s5 and s6,
s1 and s2, 
s3 and s4, respectively\,(case--I). 
Figure\,4\,(b) shows the simulated z-direction deformation, 
in which
$\overline{\alpha}_{\#1}$, 
$\overline{\alpha}_{\#2}$, and
$\overline{\alpha}_{\#3}$ 
were given for the segment 
s3 and s6,
s2 and s5, 
s1 and s4, respectively\,(case--II). 
The case--I and case--II are the configuration, in which
the two segments having the same CTE are allocated
to be adjoined and confronted position, respectively.
As the result, the configuration of the case--I
corresponds to a mirror, 
which consists of three segments having 
$\overline{\alpha}_{\#1}$,  
$\overline{\alpha}_{\#2}$,  
and 
$\overline{\alpha}_{\#3}$,
whilst the configuration of the case--II
corresponds to a mirror having a CTE distribution of
180$^{\circ}$ rotationally symmetric.
Both of figures\,4\,(a) and (b) show the surface
after the tilt correction.
For the case--I and case--II, 
0.032\,$\mu$m and 0.040\,$\mu$m\,(1\,$\sigma$ rms) 
are obtained as the surface  deformation 
after the tilt correction, respectively.
Since the requirement for the surface figure accuracy
of SPICA is 0.06\,$\mu$m rms, 
the FEM analysis indicates that the thermal deformation 
estimated based on the measured CTE dispersion among the segments 
is sufficiently small for the SPICA telescope.

The temperature expected for the SPICA is 4.5\,K,
whilst the lowest temperature of the CTE measurement in this
work is $\sim 10$\,K.
However, the thermal contraction between 10\,K and 4.5\,K is
negligible and does not affect 
the mirror deformation at all.
It is shown that the accuracy of the CTE measurement achieved 
in the present study  is sufficient to 
investigate the thermal deformation for the wave front error 
of less than 0.06\,$\mu$m for a segmented SiC mirror of 3.5m size.

\section{Conclusion}

In this work, we performed high precision 
measurement of the thermal expansion
of the sintered SiC, SiC-100 for use in  cryogenic space-telescopes. 
Three samples of SiC-100 from different lots
are measured. 
The temperature the measurement
ranges from room temperature to  $\sim 10$\,K.
The following results are obtained.

\begin{enumerate}

\item 
The typical thermal expansion of SiC-100 
is given in the form of  
$\Delta L/L = \sum_{i = 0}^8 a_i T^i$. 
The coefficients are shown in table 1.

\item 
The CTEs were measured for three samples of two orthogonal
directions.,
The average and dispersion\,(1\,$\sigma$) of these six values  
between 293\,K and 10\,K are
0.816 and 0.005\,($10^{-6}$/K), respectively.
The dispersion is well below the present measurement uncertainty.

\item
The homogeneity and the anisotropy of the CTE of SiC-100 has been
confirmed within the present measurement accuracy.

\item
The small dispersion of the absolute 
CTE obtained is compatible with the 
results of the differential CTE measurements
using brazed samples made at Astrium. 

\item
For the three samples, nominal CTEs,
$\overline{\alpha}_{\#1}$,  
$\overline{\alpha}_{\#2}$, 
and $\overline{\alpha}_{\#3}$ 
were derived for temperatures between 293\,K and 10\,K
by averaging two CTE data of tow directions for each sample.

\item 
The thermal deformation of a segmented mirror is estimated 
by a FEM analysis, using
$\overline{\alpha}_{\#1}$,  
$\overline{\alpha}_{\#2}$, 
and $\overline{\alpha}_{\#3}$. 
The result indicates that the present measured dispersion is 
sufficiently small and well below the SPICA requirement on 
thermal deformation of the mirror, 0.06\,$\mu$m rms.

\end{enumerate}

\acknowledgments
 We are grateful for Dr. Masahiro Okaji
in National Metrology Institute of Japan, 
Advance Industrial Science and Technology
for kind and large support for the measurement.
This work was supported in part by a grant from the Japan 
Science and Technology Agency.


%
%


\clearpage


\begin{figure}
\begin{center}
\includegraphics[width=10cm, angle=-90]{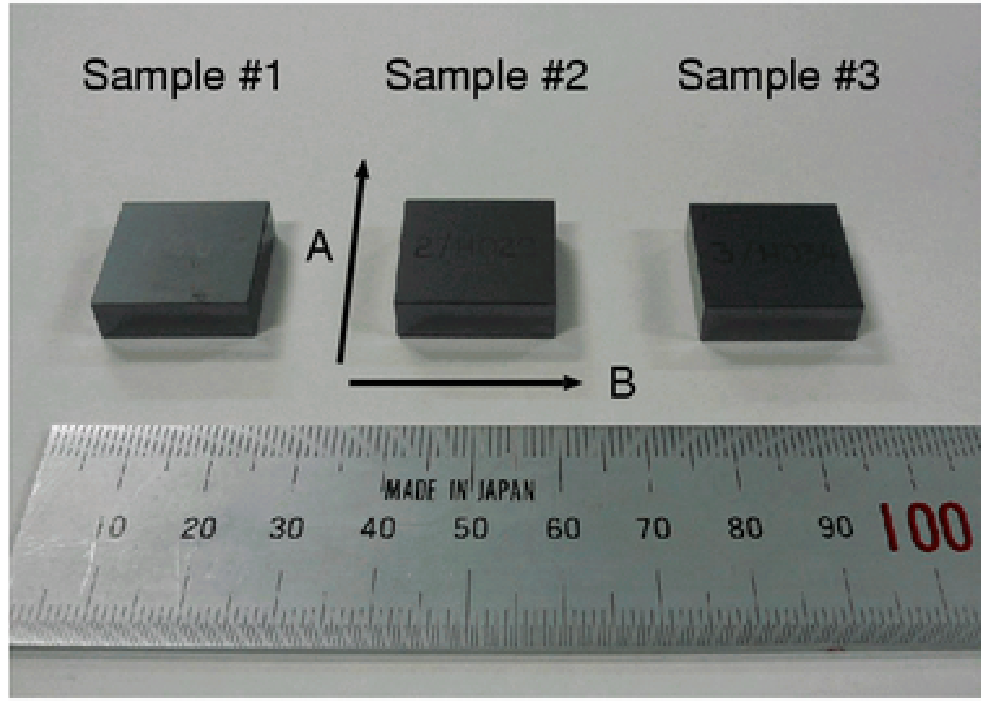}
\caption{
Samples measured in this work.
The geometry of the samples was 
20.00\,mm\,$\times$\,20.00\,mm\,$\times$\,6.0\,mm.
The surfaces of $\times 20.00$\,mm\,$\times$
6.0\,mm were polished as described in the text.
}
\label{dummy}
\end{center}
\end{figure}

\newpage

\begin{figure}
\begin{center}
x\includegraphics[width=8.8cm, angle=-90]{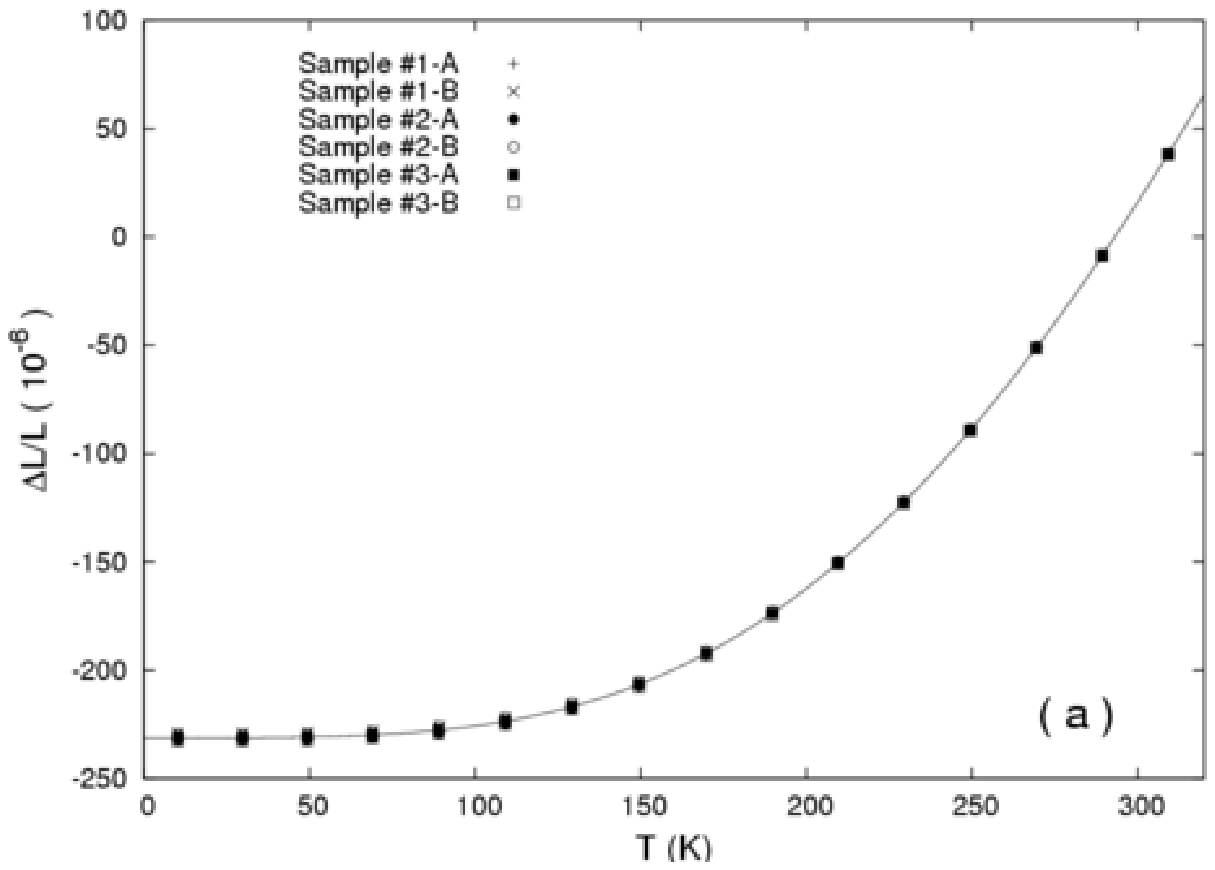}\\
\includegraphics[width=6cm, angle=-90]{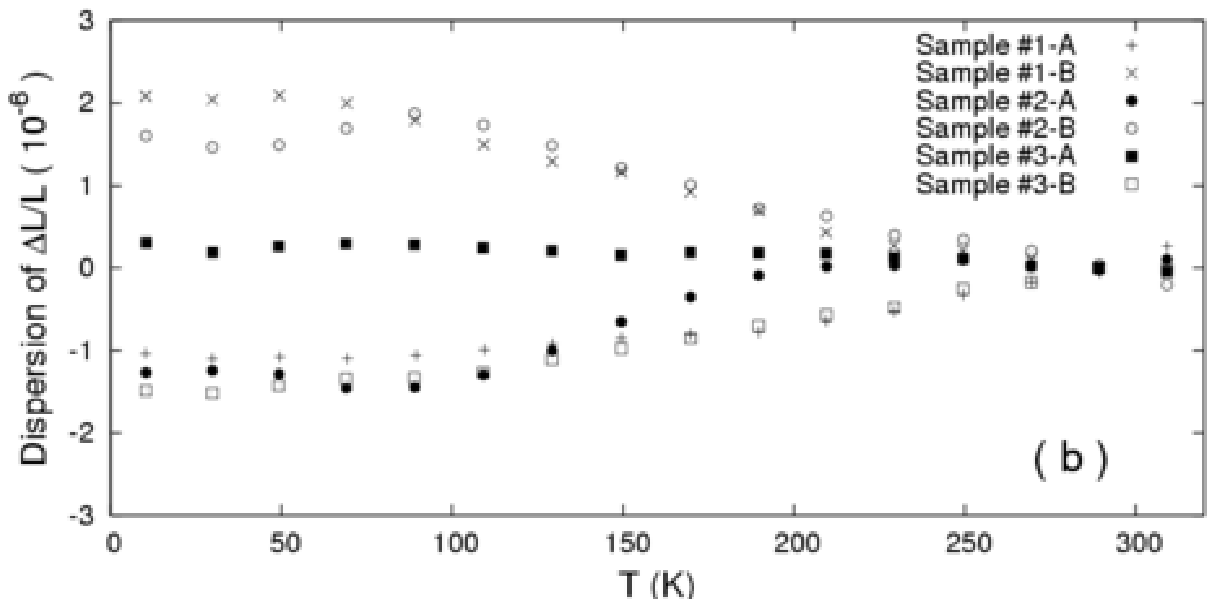}
\caption{Result of the measurements.
(a):  Thermal expansion data for 
the direction A and B of the sample \#1, \#2, and \#3.
The solid line represents the fitting curve with
8th order polynomial which is expressed by
the coefficients presented in table\,1.
(b): Residual dispersion of the thermal expansion data   
around the fitting curve.
}
\label{dummy}
\end{center}
\end{figure}

\clearpage

\begin{figure}
\begin{center}
\includegraphics[width=5.5cm,  origin=br, angle=-90]{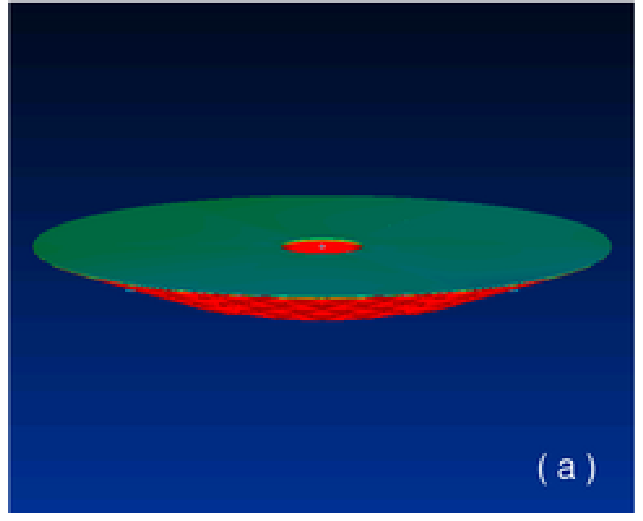}
\includegraphics[height=5.5cm]{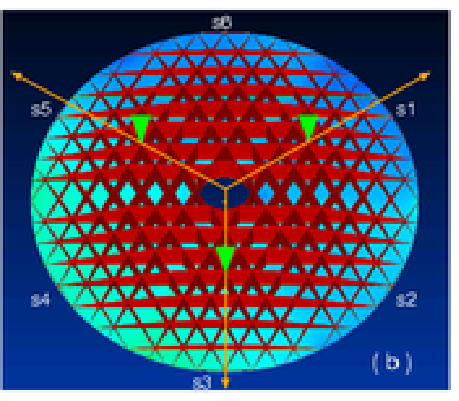}\\
\includegraphics[width=5.5cm, angle=-90]{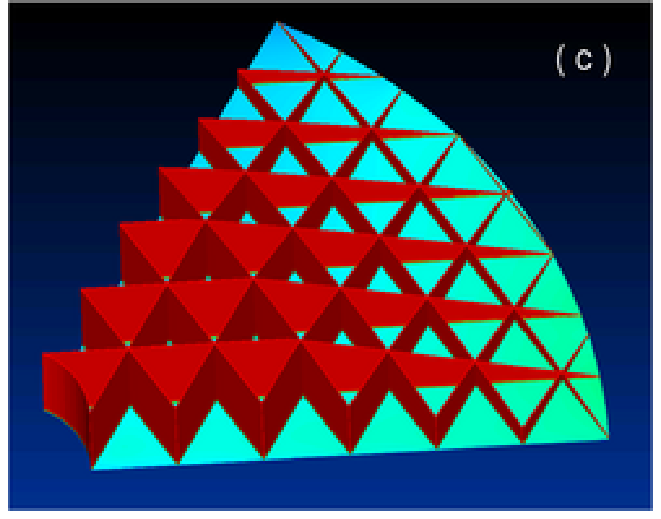}\\
\includegraphics[width=8cm, angle=-90]{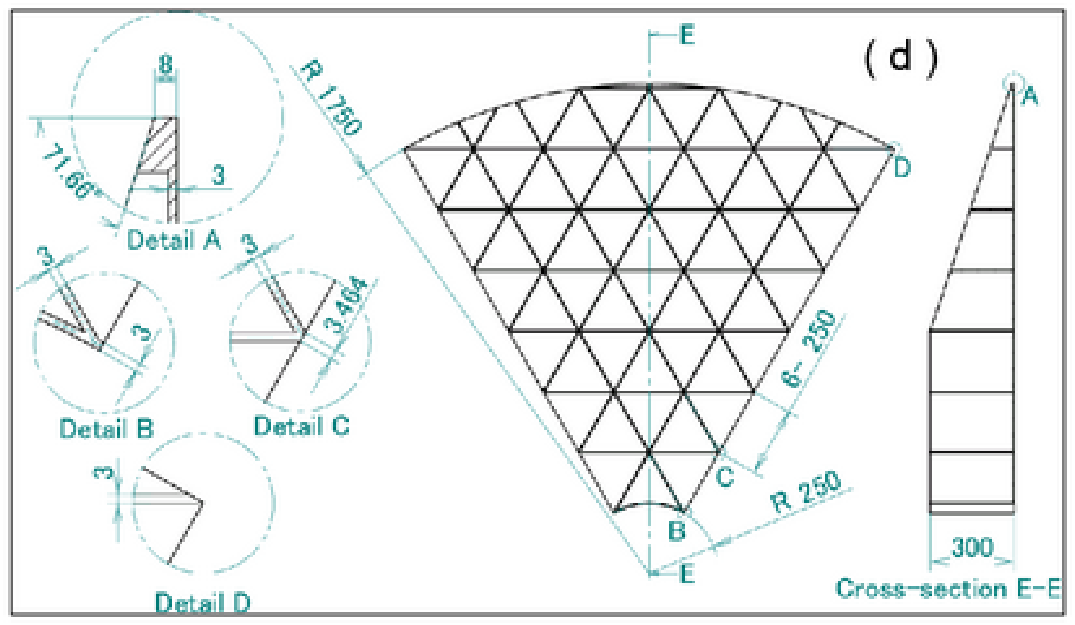}
\caption{The model for the FEM analysis. 
(a) 3\,D view of the model of the whole mirror seen 
from the reflective side.
(b) The same model but seen from the back side of the mirror.
Three triangles on the rib indicate points for constraints. 
All of the three points were constrained in the three
lines shown by the arrows. 
s1\,$\sim$\,s6 are for ID of the segments\,(see also figure\,4).
(c) and (d) shows a 3\,D view and the geometry of one segment,
respectively. 
}
\label{dummy}
\end{center}
\end{figure}

\clearpage

\begin{figure}
\begin{center}
\includegraphics[width=10cm]{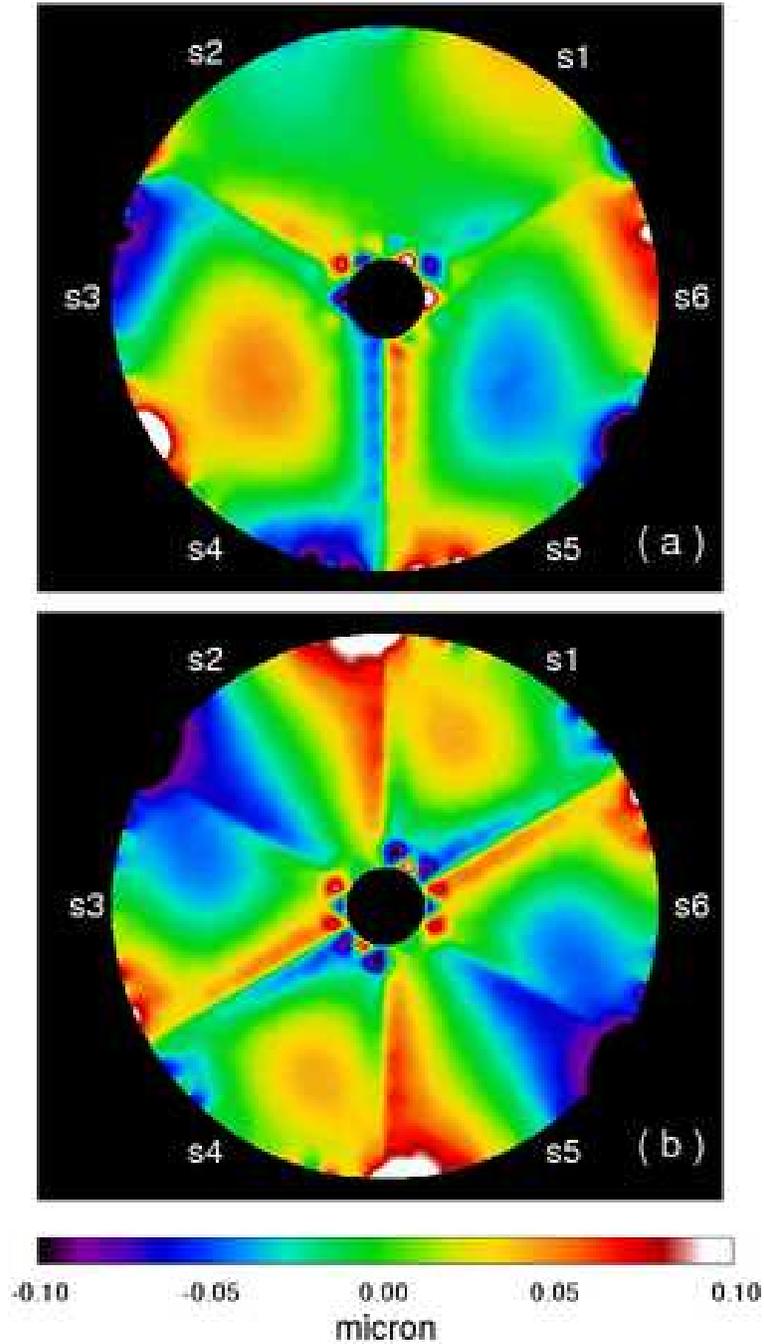}
\caption{
The simulated surface deformation by the FEM analysis. 
z-direction deformation\,(i.e., perpendicular to the surface)
after tilt correction is shown by color map.
(a): 
The result of the case--I simulation.
$\overline{\alpha}_{\#1}$, $\overline{\alpha}_{\#2}$, and
$\overline{\alpha}_{\#3}$ were given for the segment s5 and s6,
s1 and s2, s3 and s4, respectively.
(b):   
The result of the case--II simulation.
$\overline{\alpha}_{\#1}$, $\overline{\alpha}_{\#2}$, and
$\overline{\alpha}_{\#3}$ were given for the segment s3 and s6,
s2 and s5, s1 and s4, respectively. 
}
\label{dummy}
\end{center}
\end{figure}


\clearpage

\begin{table}[h]
\label{tab_8}
\begin{center}
\caption{
Coefficients of the 8th order polynomial 
$\Delta L/L = \sum_{i = 0}^8 a_i T^i$
to represent the typical
thermal expansion of SiC-100 below 300 K.
}
\begin{tabular}{cc}
\hline
\hline
   coefficient  &  value  \\
\hline
     $a_0$      & $ +2.43165 \times 10^{-1 } $ \\
     $a_1$      & $ -9.25541 \times 10^{-2 } $ \\
     $a_2$      & $ +7.38688 \times 10^{-4 } $ \\
     $a_3$      & $ +2.44225 \times 10^{-5 } $ \\
     $a_4$      & $ +5.68470 \times 10^{-7 } $ \\
     $a_5$      & $ +5.94436 \times 10^{-9 } $ \\
     $a_6$      & $ -4.04320 \times 10^{-11} $ \\
     $a_7$      & $ +8.70017 \times 10^{-14} $ \\
     $a_8$      & $ -6.64445 \times 10^{-17} $ \\
\hline
\end{tabular}
\end{center}
\end{table}

\newpage

\begin{table}[h]
\begin{center}
\caption{Summary of the measured CTE.}
\begin{tabular}{ccc}
\hline
\hline
  Sample$^{a}$  & Direction$^{b}$ 
  & $\overline{\alpha}$\,($10^{-6}/$K) $^{c}$ \\
\hline
   \#1  & A     &  0.820    \\
   \#1  & B     &  0.809    \\
   \#2  & A     &  0.821    \\
   \#2  & B     &  0.811    \\
   \#3  & A     &  0.815    \\
   \#3  & B     &  0.822    \\
\hline
\end{tabular}
\end{center}
\label{tab_six_cte}
\end{table}

a, b: \#1, \#2, and \#3 correspond to the sample number 
and A and B correspond to the direction of the sample for
measurement
shown in figure\,1. 
xc: $\Delta L/L$ between 293\,K and 10\,K.


\begin{thebibliography}{}
\bibitem[Breysse et al. 2004]{breysse}
   Breysse, J.,  Castel, D., Laviron, B., Logut, D., \& Bougoin, M., 
   2004, 
   Proc. of the 5th International Conference on Space Optics, 659
\bibitem[Enya et al. 2004]{enya2004} 
   Enya, K., Nakagawa, T.,  Kataza, H., Kaneda, H.,
   Y.Y.Yamashita, T. Onaka, T. Oshima, \& T. Ozaki,
   2004,
   Proc. of SPIE, 5487, 1092
\bibitem[Enya et al. 2006]{enya2006a}
    Enya, K.,  Nakagawa, T.,  Kaneda, H.,  Onaka, T., 
    Ozaki, T.,  \&  Kume, M.,
    2007, 
    Applied Optics, in press
\bibitem[Kaneda et al. 2003]{kaneda2003}
    Kaneda, H., Onaka, T.,  Yamashiro, R., \&  Nakagawa, T.,  
    2003,
    Proc.  of SPIE,  4850, 230


\bibitem[Kaneda et al. 2005]{kaneda2005}
     Kaneda, H.,  Onaka, T.,  Nakagawa, T., Enya, K., 
     Murakami, H.,   Yamashiro, R.,  Ezaki, T., Numao, Y., 
     \& Sugiyama, Y.,  
     2005,
     Applied Optics,  44, 6823

\bibitem[Kaneda et al. 2007]{kaneda2007}
     Kaneda, H., Kim, W., Onaka, T., Wada, T., Ita, Y., Sakon, I., \& 
     Takagi, T.,
     2007,
     Publ. Astron. Soc. Japan, submitted
     
\bibitem[Karlmann et al. 2006]{karlmann2006}
     Karlmann, P. B.,  Klein, K. J.,  Halverson, P. G.,  
     Peters, R. D.,   Levine,  M. B.,  van Buren, D., 
     \&  Dudik, M. J., 
     2006,
    Proc. of AIP Conferenc, Advances in  Cryogenic Engineering,
    824,  35


\bibitem[Murakami 2004]{murakami2004} 
    Murakami, H.,
    2004,
    Proc.  of SPIE,  5487, 330


\bibitem[Nakagawa 2004]{nakagawa2004} 
    Nakagawa, T.,  \& SPICA working group, 
    2004,
    Adv. Sp. Res.,  34, 645


\bibitem[Okaji \& Yamada 1997]{okajiyamada1997}
    Okaji, M., \& Yamada, N., 
    1997,
    High Temp. \&  High Press.,  29, 89

\bibitem[Okaji et al. 1997]{okaji1997}
    Okaji, M.,  Yamada, N.,  Kato, H., \&  Nara, H., 
    1997,
    Cryogenics, 37, 251

\bibitem[Onaka et al. 2005]{onaka2005}
   Onaka, T.,  Kaneda, H.,  Enya, K., Nakawaga, T.,  
   Murakami, H.,  Matsuhara, H.,  \&  Kataza, H.,
   2005,
   Proc.  of SPIE,  5494, 448


\bibitem[Ozaki et al. 2004]{ozaki2004} 
  Ozaki, T.,  Kume, M., Oshima, T.,  Nakagawa, T., 
  Matsumoto, T.,  Kaneda, H.,  Murakami, H., Kataza, H., 
  Enya, K.,
  T. Onaka, \& M. Krodel, 
   2004,
  Proc.  of SPIE,  5494, 366


\bibitem[Sugita et al. 2006]{sugita}
    Sugita, H.,  Nakagawa, T.,  Murakami, H.,  Okamoto,  A.,  
    Nagai, H., Murakami, H.,  Narasaki, K., \& Hirabayashi, M.,
    2006,
    Cryogenics,  46,  149


\bibitem[Toulemont 2005]{toulemont}
  Toulemont, Y.,  private communication, 2005

\bibitem[Pepi \& Altshuler 1995]{pepi1995} 
    Pepi, J. W.,  \&  Altshuler, T. L.,
    1995,
    Proc. of SPIE,  2543, 201

\bibitem[Pilbratt 2004]{pilbratt2004}  
    Pilbratt, G. T.,
    2004,
    Proc.  of SPIE,  5487, 401


\bibitem[Suyama et al. 2005]{suyama2005} 
     Suyama, S.,  Itoh, Y.,  Tsuno, K., \&  Ohno, K., 
     2005,
     Proc.  of SPIE,  5868, 96


\bibitem[Yamada \& Okaji 2000]{yamada2000}
    Yamada, N., \&  Okaji, M.,
    2000, 
     High Temp. \& High Press., 32, 199




\end{thebibliography}
\end{document}